\documentclass[%
 preprint, 
 amsmath,amssymb,
 aps, 
rmp,
]{revtex4-2}

\usepackage{graphicx}
\usepackage{bm}
\usepackage{amsmath}
\usepackage{xcolor}
\usepackage{tabularray, lipsum}
\usepackage{soul}

\begin{document}

\preprint{APS/123-QED}

\title{\textbf{Isotopes for Medicine and Fundamental Physics} 
}%

\author{V. J. Ajith}

\author{Mark G. Raizen}%
 \email{Contact author: raizen@physics.utexas.edu}
\affiliation{%
 Authors' affiliations\\
  Department of Physics, The University of Texas at Austin, Austin, Texas, USA.
}%




\date{\today}

\begin{abstract}

Isotopes, variants of elements distinguished by their neutron number, have important applications for medicine and fundamental science. This review explores the historical challenges and new technologies for enriching and detecting isotopes. We motivate the work by challenges posed by three of history's greatest scientists: James Clerk Maxwell's thought experiment on thermodynamics, Ernest Rutherford's practical dilemma in radiochemistry, and Albert Einstein's questions about the probabilistic nature of quantum mechanics. The journey to meet these challenges has catalyzed the development of innovative techniques for isotope separation. These advancements can mitigate the worldwide crisis in the supply of enriched isotopes. We then explore the expected impact of these new technologies on medicine, which include radioisotopes for cancer therapy and stable isotopes as tracers for metabolism. We explain the importance of developing scalable techniques for isotope detection and how it can help with early detection of certain diseases. Finally, we turn to basic science, outlining proposed new tests of quantum mechanics that utilize radioactive atomic clocks to probe the fundamental nature of determinism and distinguishability of atoms. 

\end{abstract}

\maketitle

\tableofcontents

\section{\label{sec1}Background and Motivation}

The study of the physical world is often propelled by challenges, questions posed by brilliant minds that push the boundaries of theory and experiment for generations. This article is framed around three such seminal challenges from James Clerk Maxwell, Ernest Rutherford, and Albert Einstein. The common thread connecting these disparate scientific quests, spanning thermodynamics, nuclear physics, and quantum mechanics, is the isotope. An isotope is a variant of a particular chemical element; while all atoms of an element have the same number of protons in their nucleus, isotopes of that element have different numbers of neutrons. For instance, the element carbon is defined by its 6 protons. The most common isotope, carbon-12 ($^{12}$C), has 6 neutrons, while the rarer, stable carbon-13 ($^{13}$C) has 7 neutrons. Because the number of electrons (which governs chemical behavior) is determined by the number of protons, isotopes of an element have nearly identical chemical properties and very similar physical properties, making them extraordinarily difficult to separate from one another.

Isotopes find applications in industry, energy production, cancer treatments, and physics experiments. The 2009 report from the Nuclear Science Advisory Committee lists many of these applications\cite{nsac_isotopes_subcommittee_isotopes_2009}. There are good reviews on different treatments with radiopharmaceuticals and imaging techniques with radioisotopes\cite{zhang_radiopharmaceuticals_2025}. This colloquium review complements the existing literature by informing the community about the potential impact of new isotope enrichment technologies like Magnetically Activated and Guided Isotope Separation (MAGIS) and Multi Resonant Laser Isotope Separation (MRLIS) in lowering isotope costs and improving isotope accessibility. We highlight the use of MRLIS to avoid radiochemistry and open a direct route to producing different radioisotopes used in treating cancer, including lutetium-177, which is in high demand. Using isotopes in medicine is not just limited to radioisotopes. In this review, we cover the use of stable isotopes in medicine. Stable isotope ratios are used as metabolic tracers to address global issues like iron deficiency. Stable isotope metalomics shows promise for early detection of various diseases, including cancers. We identified that, along with the availability of isotopes, detecting isotopes is also important and needs innovation. We discuss the use of laser isotope spectroscopy to detect isotopes in biological samples. Then, in the application of isotopes in fundamental physics, we highlight newly proposed experiments with trapped radioactive ions. We start the review by giving a historical background to our understanding of isotopes. 

\subsection{\label{sec1_1}The Discovery and Conception of Isotopes}
The notion that atoms of a single element could have different masses was a revolutionary concept that emerged from the study of radioactivity in the early twentieth century. Researchers investigating the decay chains of uranium and thorium discovered numerous new radioactive substances, far more than there were available spots in the periodic table \cite{thoennessen_discovery_2016}. In 1913, the British chemist Frederick Soddy \cite{soddy_1913}, upon realizing that chemically inseparable elements could have different radioactive properties and atomic weights, proposed the concept of "isotopes," from the Greek words \textit{isos}("same") and \textit{topos} ("place"), because these elemental variants occupied the same place in the periodic table.

Concurrently with these radiochemical discoveries, the first direct physical evidence for isotopes in a stable element came from the laboratory of J.J. Thomson \cite{thomson_bakerian_1913}. In 1912, while studying streams of positive ions ("canal rays"), Thomson deflected a beam of neon ions using parallel magnetic and electric fields. The ions traced parabolic paths on a photographic plate, with the exact path determined by their charge-to-mass ratio. He observed two distinct parabolas, corresponding to atoms with masses of approximately 20 and 22 atomic mass units, providing compelling evidence that the stable element neon was a mixture of at least two different species \cite{thomson_bakerian_1913}. This work was further refined by Thomson's student, Francis William Aston, who developed the first true mass spectrograph. With this instrument, Aston could precisely measure atomic masses and discovered that the masses of the isotopes of non-radioactive elements were, to a close approximation, integer multiples of the mass of a hydrogen atom—his "whole number rule" \cite{aston_bakerian_1927}. This work systematically revealed that most elements are mixtures of isotopes and earned Aston the 1922 Nobel Prize in Chemistry.


\subsection{\label{sec1_2}The Challenge and Methods of Isotope Separation}

The discovery of isotopes immediately presented a formidable technical challenge: how to separate them. Since isotopes of an element are chemically identical, standard chemical techniques are ineffective, except for the lightest elements. Separation must rely on the small physical differences between them, primarily their mass. During the twentieth century, several large-scale industrial methods were developed, largely driven by the urgent need to enrich uranium in its fissile isotope, uranium-235 ($^{235}$U), for nuclear applications. There are different kinds of isotope separation methods like Electromagnetic Isotope Separation (EMIS), Gaseous Diffusion, Gas Centrifuge, and Laser Isotope Separation (LIS).

Electromagnetic Isotope Separation (EMIS), embodied by the Calutron, is essentially a scaled-up version of Aston's mass spectrograph. The process involves ionizing a sample, accelerating the ions into a magnetic field, and collecting them based on the differing radii of their curved trajectories. Developed by Ernest Lawrence at Berkeley during the Manhattan Project, the Calutron was a monumental undertaking. The Y-12 plant at Oak Ridge, Tennessee, housed over a thousand Calutrons operating in massive arrays, or "racetracks" \cite{love_electromagnetic_1973}. While it can achieve very high purity in a single pass and is versatile enough to separate isotopes of almost any element, the Calutron's primary drawback is its extreme inefficiency. The process suffers from low throughput and space-charge effects that limit beam intensity \cite{alexeff_instability_1978}. Most critically, it is incredibly energy-intensive, requiring enormous electromagnets to maintain fields of up to 1 Tesla over volumes of many cubic meters. This translates directly into the prohibitively high cost of enriched isotopes today, for instance, 10 mg of iron-58 costing around \$3,500\footnote{Isotope quotation for iron-58 obtained from National Isotope Development Center, Oak Ridge National Laboratory(ORNL) on October 8, 2025}.

Gaseous diffusion was the primary production method for enriching uranium for many decades. It relies on Graham's law of effusion, which states that the rate at which a gas passes through a porous barrier is inversely proportional to the square root of its mass \cite{sherr_separation_1938}. To apply this to uranium, it is first converted into a gaseous compound, uranium hexafluoride ($\text{UF}_6$). When this gas is pumped against a barrier with microscopic pores, the lighter $^{235}\text{UF}_6$ molecules pass through slightly faster than the heavier $^{238}\text{UF}_6$ molecules. The separation factor in a single stage is very small (about 1.0043), so achieving significant enrichment requires cascading thousands of these stages together in enormous facilities, making this process even more energy-intensive than the Calutron.

The Gas centrifuge method has now largely supplanted gaseous diffusion for uranium enrichment due to its superior energy efficiency (requiring only about 4-5\% of the electricity of a diffusion plant for the same output \cite{olander_gas_1978}). As demonstrated in early work by Jesse Beams \cite{beams_separation_1936,beams_high_1938}, a gas centrifuge is a rapidly spinning cylinder containing $\text{UF}_6$ gas. The immense centrifugal force (many thousands of times the force of gravity) moves the heavier $^{238}\text{UF}_6$ molecules toward the cylinder wall, while the inner region becomes slightly enriched in the lighter $^{235}\text{UF}_6$. By inducing a countercurrent flow within the cylinder, this small radial separation can be converted into a larger axial separation, and by connecting many centrifuges in a cascade, high levels of enrichment can be achieved. The separation factor of earlier systems is around 1.25 for a single stage \cite{olander_gas_1978}. This method is economical for isotopes that have gaseous compounds near room temperature. But many of the medically relevant isotopes do not satisfy this condition. Moreover, radioisotopes needed for targeted therapy, which we will discuss in Section \ref{sec3}, require higher enrichment than this technique can provide.

Laser Isotope Separation (LIS) is a relatively newer approach that offers the promise of much higher selectivity and efficiency by exploiting the subtle differences in the electronic energy levels between isotopes. The isotope shift causes different isotopes to absorb light at slightly different frequencies. In Atomic Vapor Laser Isotope Separation (AVLIS) \cite{paisner_atomic_1988, solarz_physics_1985,yakovlenko_main_1998}, a laser is precisely tuned to selectively excite and ionize only the atoms of the desired isotope in a vapor. These newly created ions can then be easily separated from the neutral atoms of other isotopes using an electric field. AVLIS can provide a much higher throughput than the Calutron and with much better energy efficiency. The drawbacks of AVLIS are the need for high power lasers which suffer from reliability issues, low efficiency of ionization, in the range of 3-5\%, and enrichment that is comparable to the Calutron. 

The two methods developed in our group, MAGIS and MRLIS, are also laser-based and have several key advantages compared with AVLIS. The first method we developed, MAGIS \cite{raizen_magnetically_2012,mazur_demonstration_2014, raizen_isotope_2014}, was inspired by the one-way wall and Maxwell’s demon, which we will see in the next section. In Section \ref{sec2_4} we explore MAGIS in detail and compare it with other separation techniques. A very recent method, MRLIS \cite{barr_efficient_2025}, was proposed as an alternative to MAGIS, and experimental demonstration is in progress in our group. MRLIS relies on multi-photon ionization, like AVLIS, but utilizes optical pumping with continuous-wave, low-power lasers. Numerical simulations predict very high degrees of enrichment, near-unity isotope capture efficiency, and a footprint that can easily fit inside a hot cell of an irradiation facility that generates radioisotopes. We believe that MRLIS will be very effective in separating radioisotopes as a replacement for radiochemistry. It can enable separation when radiochemistry is impossible, such as the production of lutetium-177($^{177}$Lu) from lutetium-176($^{176}$Lu). We will discuss this further in Section \ref{sec3}. As a convenient reference, we summarize different isotope-separating techniques in table \ref{table1}.

\begin{longtblr}[
caption = {Comparison of different isotope separation techniques},
  label = {table1}
                ]{colspec={X[l] X[l] X[l] X[1,l] X[2,l] X[l] X[1,l]},
                  rowhead = 1,
                  row{1} = {c, font=\bfseries},
                hlines, vlines 
                }
&Through-put	&Isotope purity	&Energy requirement	&Limitation	&Nuclear isomer selection	&Suitable for Radioisotopes in Medicine
         \\
Gas diffusion	&High	&Low	&High	&Need gaseous compounds of isotopes near room temperature	&No	&No
 \\
 Gas centrifuge& High& Low& Low& Need gaseous compounds of isotopes near room temperature& No& No\\
Calutron& High& High& High& High energy demand& No& No\\
MAGIS& Low& Highest& Lowest& Use of permanent magnet limit the magnetic field strength causing long isotope filtering tracks& Yes& No\\
AVLIS& Moderate to high& High& Low& Duty cycle and linewidth of the pulsed laser limit the throughput and purity of the isotope& Yes& No\\
MRLIS& Moderate to high& High (Higher than AVLIS) & Low (Lower than AVLIS)& Needs a high mode volume laser cavity which needs precision alignments.& Yes& Yes (Compact to fit in a hot cell)\\

\end{longtblr}


\subsection{\label{sec1_3}The Modern Isotope Supply Crisis and its Impact}

Despite the existence of these technologies, the world faces a growing crisis in the supply of enriched stable isotopes \cite{cutler_global_2021}. Russia(85\%) and Netherlands(15\%) are the main suppliers of large quantity market of stable isotopes \cite{nsac_isotopes_subcommittee_meeting_2015}. Large quantity market involves the isotope sale of a few thousand kilograms per year. Medically relevant precursor isotopes like $^{203}$Tl, $^{191}$Ir, $^{88}$Sr, and $^{68}$Zn come under this large quantity market. Gas centrifuge is the main technique used in the large quantity market, but as we have seen centrifuge is limited to only a few isotopes. Small quantity market for the research and medical applications is supported by the stable isotope reserves from Oak Ridge National Laboratory (ORNL) in US and from Russia. Most of the US isotope reserves are from the Calutrons in ORNL. Since the shut down of the Calutrons in 1998, the isotope reserves are slowly depleting. This has left the U.S. and the world heavily reliant on a few aging international facilities, primarily in Russia \cite{nsac_isotopes_subcommittee_isotopes_2009,nsac_isotopes_subcommittee_meeting_2015}. 

Stable isotopes are the precursors in the production of radioisotopes. As we will see in Section \ref{sec3}, they are used in cancer treatment. The stable isotopes undergo a nuclear reaction in a cyclotron or a nuclear reactor to produce the required radioisotope. The production of radioisotopes often needs to be close to the diagnostic or treatment center because of the short half-life of radioisotopes. There are a few of companies who make the radioisotopes and make the radio-labelled drug(radiopharmaceutical) from the radioisotope. In the US, research reactors in universities are also used for the local production of the radio-isotopes\cite{nuclear_energy_agency_supply_2010,cutler_global_2021}. Reliance on these limited number of facilities, compounding to the availability of stable precursor isotopes, makes the supply chain of radioisotopes precarious. We will discuss more specific challenges of radiopharmaceutical production in Section \ref{sec3_2}.

This fragility in the supply chain has profound consequences. The high cost and lack of availability of enriched isotopes create a significant bottleneck for research and development in medicine, national security, and basic science. Medical therapies that require precursor isotopes for their production face uncertain supply and volatile pricing \cite{cutler_global_2021}. Scientific experiments that could push the frontiers of physics may be delayed or deemed infeasible due to the prohibitive cost of their isotopic targets. This situation has created an urgent, universally recognized need for new, cost-effective, and energy-efficient methods of isotope separation. The technologies described in this review, born from fundamental physics, are poised to meet this critical national and global need.


\section{\label{sec2}Controlling Atoms with Maxwell’s Demon}

\subsection{\label{sec2_1}Maxwell's Challenge and the Thermodynamic Debate}

In 1871, James Clerk Maxwell proposed a thought experiment that would challenge the very foundations of thermodynamics. Maxwell is, of course, best known for the so-called Maxwell’s equations, which unified electricity, magnetism, and light. But towards the end of his life, Maxwell came up with a thought experiment, known as Maxwell’s demon; he considered a chamber of gas-phase molecules divided into two compartments, ‘A’ and ‘B’, by a partition containing a tiny trapdoor. If it’s in the closed position, molecules inside ‘A’ and ‘B’ cannot go back and forth. Molecules can move across only when the trap door is open. This trapdoor is controlled by a creature known as the demon, who can see the coming and going of every gas-phase molecule. For example, if the demon sees a molecule in chamber ‘B’ moving towards the trapdoor, the demon could open it and let the molecule pass through. Otherwise, the demon keeps it closed. And in such a manner, the demon could cause all the molecules from ‘B’ to move into ‘A’ and none of the molecules from ‘A’ to move into ‘B’, thereby concentrating everything inside ‘A’ without altering the kinetic energy of the particles. This seemed to violate the second law of thermodynamics because the system became more ordered by increasing the density without doing work or changing the temperature. Thinking further, if all molecules are concentrated in ‘A’, then, keeping the trapdoor closed, expanding that wall adiabatically would cause the particles to cool down as their kinetic energy is reduced. Repeating these steps will then reduce the temperature of the gas-phase molecules. This thought experiment has been the topic of debate for almost 140 years without a practical resolution except for single-particle demonstration experiments \cite{maruyama_colloquium_2009}. To summarize Maxwell’s challenge from 1871, it was to realize Maxwell’s demon with large ensembles.

For decades, this paradox sparked intense debate. An early and crucial physical argument against the demon came from Polish physicist Marian von Smoluchowski in 1912 \cite{smoluchowski_experimental_1912,maruyama_colloquium_2009}. He pointed out that any real, physical demon and its trapdoor mechanism would themselves be subject to the thermal environment. The random bombardment of molecules (Brownian motion) would cause the trapdoor to fluctuate uncontrollably, rendering the demon's sorting attempts futile. To operate reliably, the demon would need to be at a lower temperature than the gas it was sorting, which simply shifts the problem.

A more elegant illustration of the second law's resilience was provided by Richard Feynman in his famous "ratchet and pawl" thought experiment \cite{feynman_feynman_2011}. He imagined a tiny paddle wheel connected by an axle to a ratchet. Gas molecules striking the paddles from one side would cause it to turn, and the ratchet would ensure it could only turn in one direction, seemingly allowing it to do work (e.g., lift a weight) by harnessing random thermal motion. However, Feynman showed that if the entire system, including the pawl and ratchet, is at the same temperature as the gas, the pawl itself will jiggle randomly due to thermal fluctuations. It will occasionally bounce up at just the right moment to allow the ratchet to turn backward, exactly canceling out any net forward motion. The only way for the device to do net work is if the pawl is kept colder than the paddle wheel, again demonstrating that a temperature difference is required to extract useful work from thermal energy.


\subsection{\label{sec2_2}Information Theory and the Resolution of the Paradox}

The modern resolution of Maxwell's paradox came not from mechanics, but from the field of information theory. In 1929, Leó Szilárd first formally linked the demon's action to the acquisition of information \cite{maruyama_colloquium_2009,leff_maxwells_2002}. He argued that for the demon to know which side of a partition a molecule is on, it must first perform a measurement. This act of "finding out" has a thermodynamic cost. This line of reasoning culminated in a principle articulated by Rolf Landauer at IBM in 1961 \cite{landauer_irreversibility_1961}. Landauer's principle states that any logically irreversible manipulation of information, such as the erasure of a bit from a computer's memory, is necessarily accompanied by the dissipation of a corresponding amount of energy into the environment, increasing its entropy.

The minimum entropy increase associated with erasing one bit of information is $k_B ln(2)$, where $k_B$ is the Boltzmann constant. In the context of Maxwell's demon, the demon's brain or memory fills up with information about the molecules it has sorted. To continue sorting, it must periodically erase this memory to make room for new information. This act of erasure generates entropy in the demon's own system or its immediate environment. It was shown that this entropy increase is always greater than or equal to the entropy decrease achieved by sorting the gas, thus perfectly preserving the second law of thermodynamics. The paradox was resolved: the demon doesn't get a "free lunch"; it pays for the ordering of the gas with the currency of information.

These theoretical models did not really explain how this would be implemented. Whether such a demon could be realized in practice remained 
unclear. Certainly, one could imagine demonstration experiments on a single particle that could realize the demon \cite{toyabe_experimental_2010}, but Maxwell’s challenge really dealt with large ensembles. And it is fair to say that even today, the idea that we could detect and then act to remove the kinetic energy, or to concentrate molecules this way, still seems impossible to us. So, instead of addressing this challenge head-on, we presented a model that seems, at first sight, unrelated but turns out to be the key to this question. Suppose we have a box of molecules, and we put a wall in between, but we make this wall so that the atoms or molecules can go one way but cannot go back. We next show that the construction of such a self-acting device is indeed possible in practice.


\subsection{\label{sec2_3}From a Thought Experiment to a Physical Device: The One-Way Wall for Atoms}

While information theory provided a satisfactory theoretical resolution, the question of whether a physical, self-acting device could achieve the demon's goal remained. The breakthrough came from the realization that such a device could be constructed by combining a conservative potential and an irreversible process. A concrete proposal for such a "one-way wall" for atoms was made in 2005 \cite{raizen_compression_2005} and experimentally demonstrated in 2008 \cite{price_single-photon_2008, thorn_experimental_2008}.

The key irreversible process is optical pumping, a technique pioneered by Alfred Kastler for which he received the 1966 Nobel Prize in Physics. This process uses laser light to manipulate the internal quantum states of atoms. The interaction of an atom with an external magnetic field is governed by its magnetic quantum number, \textit{m}. The energy levels of an atom in a magnetic field split into multiple sublevels (the Zeeman effect), and the force on the atom depends on which sublevel it occupies. Atoms in an \textit{m}=0 state are largely unaffected by a magnetic field, while those in other states (e.g., \textit{m}=1) can be strongly repelled or attracted. Optical pumping allows physicists to use laser light of a specific frequency and polarization to move the entire population of atoms into a chosen \textit{m} state. This process is irreversible because it involves the spontaneous emission of photons, which are radiated in random directions.

The one-way wall operates as follows:
\begin{enumerate}
    \item A cloud of atoms is confined in a box and initially prepared in a non-magnetic \textit{m}=0 state.
    \item A magnetic field, acting as a potential barrier, is established in the center of the box. The atoms, being in the \textit{m}=0 state, can pass through this barrier freely.
    \item A laser beam is introduced on one side of the barrier (say, the left side). This laser is tuned to optically pump any atom passing through it from the non-magnetic \textit{m}=0 state to a magnetically repulsive \textit{m}=1 state.
\end{enumerate}

\begin{figure*}
\includegraphics[width=1.0\textwidth]{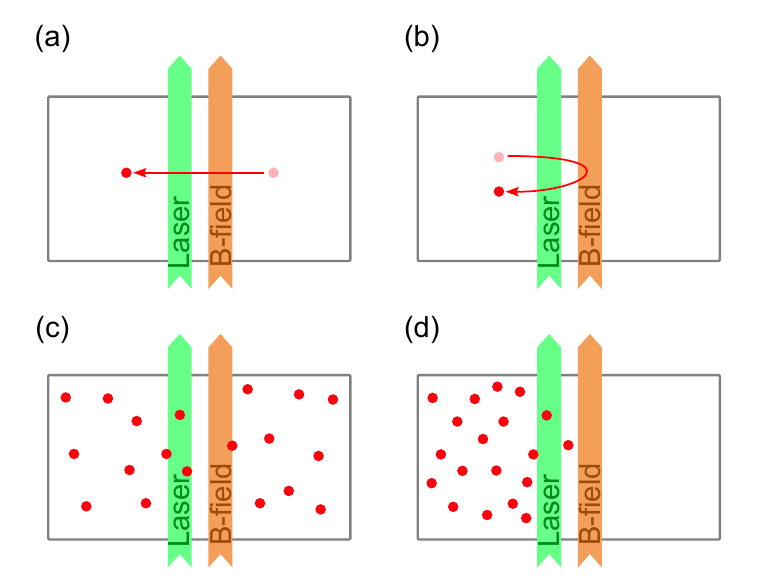}
\caption{\label{fig1} One way wall for atoms made with a laser beam and magnetic field. The laser can excite an atom to a magnetically repelling state. (a) An atom moving from right to left goes through the wall. Here the excitation happens after the atom crosses the magnetic field. (b) An atom moving from left to right get excited by the laser first and magnetic field repels the atoms. (c) initial state of a chamber with a few atoms with a laser beam and magnetic field. (d) final state where all atoms gather on the left side. }
\end{figure*}

The result is a one-way wall \cite{raizen_compression_2005}. An atom from the right side can cross the barrier (FIG. \ref{fig1}(a)), enter the left side, get "state-flipped" by the laser, and then become trapped on the left side because it is now repelled by the magnetic barrier. Any atom on the left side is also optically pumped and subsequently repelled by the barrier (FIG. \ref{fig1}(b)), preventing it from leaving. Over time, the entire population of atoms accumulates on the left side (FIG. \ref{fig1}(d)), compressing and cooling the gas without a physical piston. This represents a significant decrease in the entropy of the atomic gas or compression of phase space. In subsequent work, we observed a 350 times compression of phase space \cite{travis_bannerman_single-photon_2009}, which is a significant factor.

This method demonstrated that this one-way wall, in fact, works, and it can reduce the entropy of gas. It was not a perfect method because it did not conserve atom number; atoms were lost in the process, and since then, we have come up with an alternative approach, which does conserve atom number and should be very efficient, and this was published in 2023 \cite{hillberry_efficient_2023} though the experiment still has not yet been done. This is potentially a way to cool atoms further when they are confined in a magnetic trap, so this could be an alternative to evaporative cooling of atoms; it could be applied to produce an ultra-bright atom laser and could be useful for atom interferometry for quantum-limited inertial sensing.

But coming back to the one-way wall for atoms, the self-acting device, is this Maxwell’s demon? And so, it’s interesting to ask what would Maxwell say?  In a letter that he wrote to John William Strutt (Lord Rayleigh) on December 6th of 1870 \cite{maxwell_scientific_1990} Maxwell said, "I do not see why even intelligence might not be dispensed with and the thing made self-acting." So, a self-acting device is what Maxwell envisioned, and that is exactly what the one-way wall does. So, in all respects, it is Maxwell’s demon. It also answers the question, what is happening with the entropy; as we showed in another paper, the entropy of the gas goes down, but we can account for that by the increase of entropy of the scattered photons from the laser into free space, thereby increasing the photon entropy \cite{ruschhaupt_one-photon_2006}. So, one does not need advanced knowledge, which was the pictorial representation of Maxwell’s demon and was a stumbling block, clearly impossible for large ensembles. Here it’s a self-acting device, but it resolves all questions about Maxwell’s demon, and most importantly, it works in practice on large ensembles.  And so, we can say we’ve met the challenge of Maxwell’s demon. We proposed a few years ago that a similar construction can be effectively used for isotope separation.


\subsection{\label{sec2_4}Magnetically Activated and Guided Isotope Separation (MAGIS)}

The principles of the one-way wall—using light to change an atom's state and a magnetic field to alter its path—can be adapted into a highly efficient isotope separation scheme. This method, known as Magnetically Activated and Guided Isotope Separation (MAGIS), was proposed as an energy-efficient replacement for the Calutron \cite{raizen_magnetically_2012,mazur_demonstration_2014, raizen_isotope_2014}.

The MAGIS apparatus begins with an atomic beam generated from a heated crucible in a vacuum. This beam is directed towards a curved guide structure made of permanent magnets arranged in a Halbach array; a special arrangement of magnets that creates a very strong field on one side while nearly canceling the field on the other, all without any power consumption. The guide is curved to block any direct line-of-sight path from the source to the collector.

\begin{figure*}
\includegraphics[width=1.0\textwidth]{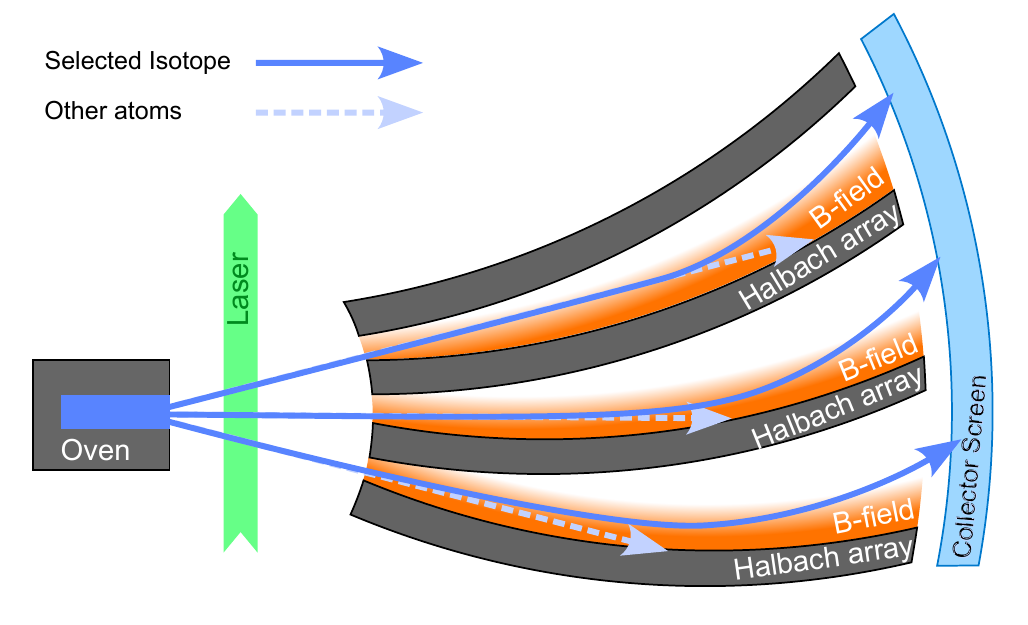}
\caption{\label{fig2} : Schematic figure for Magnetically Activated and Guided Isotope Separation (MAGIS). Different isotopes of an element have slightly different excitation frequency (isotope shift). A laser with narrow linewidth is tuned to excite atoms of a particular isotope to a magnetically repelling state. Atoms then enter a stack of curved guides with magnetic fields in each guide due to a Halbach array of permanent magnets. The isotope that is repelled get collected on a screen (solid arrows from the oven) at the end of the guide while other atoms collect on the sides of the guide (dashed arrows from the oven). }
\end{figure*}

The separation mechanism leverages the isotope shift. A schematic figure of the MAGIS is shown in FIG. \ref{fig2} with multiple curved magnetic guides. An isotopically selective laser illuminates the atomic beam before it reaches the magnetic guide. The laser is tuned to excite and optically pump only the desired isotope into a magnetically repulsive state (a "low-field-seeking" state). Depending on the isotope abundance and which isotope is being enriched, different strategies can be used to in the selective optical pumping \cite{raizen_magnetically_2012}. The unwanted isotopes remain in their default state, which is either unaffected by or attracted to the magnetic field. When the beam strikes the guide, the desired, repelled isotopes are deflected by the magnetic field and guided around the curve to a collection plate. The unwanted isotopes are not deflected, follow a straight path, impact the magnet array liner, and are removed.

This process is directly analogous to Maxwell's original conception of a "pointsman on a railway," intelligently switching tracks to sort different types of trains.

In a letter to Lord Kelvin in January 16$^{th}$ of 1868 \cite{collins_kelvin_2015}, he said; “now let an intelligent but finite and energy-conserving being be appointed pointsman to the flying molecules”. In another letter, in December 6$^{th}$ of 1870 \cite{maxwell_scientific_1990}, he wrote to Lord Rayleigh, “like a pointsman on a railway with perfectly acting switches, who should send the express along one line and the goods along another”. Maxwell was referring to the pointsman of the railways. This was a person in the 19$^{th}$-century British Empire who was responsible for switching of the tracks, so this is a very important responsibility because it ensures the safety of the railway system. Maxwell was intrigued by the pointsman because it was an example in his mind of the exertion of free will, as opposed to Newtonian physics, where everything is determined by the equations of motion. Once one knows the initial conditions and the interaction, or the potential, future evolution is determined. So, Maxwell was inspired by the pointsman of the railway, and in fact, this was the origin of Maxwell’s demon. He envisioned a stream of atoms where you switch the tracks, not the box with the gas phase particles; that is very reminiscent of the MAGIS separation process. So, in a very real way, MAGIS is the closest to Maxwell’s idea of the pointsman, which is the precursor to Maxwell’s demon. Indeed, this analogy was the inspiration for the non-profit Pointsman Foundation (www.pointsman.org), created to leverage MAGIS technology for producing affordable medical isotopes. 

The advantages of MAGIS over Calutron are manifold. It is vastly more energy-efficient, as it relies on permanent magnets and low-power lasers. The main electrical consumption is to vaporize the element inside an oven to form an atomic beam, and to operate the vacuum pumps. The lasers are low-power and solid-state, so their energy consumption is negligible. Its selectivity is also superior. The Calutron separates based on the tiny fractional difference in the charge-to-mass ratio (q/\textit{m}). MAGIS separates based on the magnetic-moment-to-mass ratio ($\mu$/\textit{m}). Optical pumping can change $\mu$ dramatically—making it zero or even flipping its sign—creating a much larger and more robust separation effect. The enrichment with MAGIS is extremely high, demonstrated first with lithium isotopes to a level of 99.95\%\cite{mazur_demonstration_2014}, and later with an isotope of ytterbium to a level of 99.98\%. The latter case, unpublished from a commercial machine, suppressed unwanted isotopes by a factor of over 50,000, unmatched anywhere. This degree of separation is necessary for medically relevant isotope separation were purity is important. The drawback of MAGIS relative to AVLIS is that the isotope flux is lower, but that is unimportant for medically relevant stable isotopes where the quantities are very small. Some of the facts discussed here were also used in the table \ref{table1}. Ongoing work on next-generation methods like MRLIS promises to further refine these laser-based techniques \cite{barr_efficient_2025}, potentially allowing for compact systems that can operate inside shielded "hot cells" for separating radioactive isotopes, for one of the most important applications of isotope which is cancer treatment. In the next section, we will briefly introduce isotope requirements in cancer therapy and Rutherford's challenge. We also project the use of MAGIS and MRLIS to supply precursor stable isotopes in the early stages of radiopharmaceutical production, and the use of MRLIS again in the purification of radioisotopes in the later stages of radiopharmaceutical production. 


\section{\label{sec3}Cancer Therapy}

The ability to efficiently separate isotopes will promote advances in medicine, particularly in the fight against cancer. According to the World Health Organization(WHO), cancer is a leading cause of death worldwide, accounting for nearly one in six deaths. It is a group of diseases characterized by the uncontrolled proliferation of abnormal cells that can invade and destroy normal body tissue. Radioisotopes have emerged as a cornerstone of modern oncology, providing powerful tools for both diagnosis and therapy \cite{zhang_radiopharmaceuticals_2025}.

\subsection{\label{sec3_1}Targeted Cancer Therapy with Radioisotopes}

The field of nuclear medicine leverages the unique properties of specific radioisotopes to target and destroy cancer cells while minimizing damage to surrounding healthy tissue. This approach, known as targeted radionuclide therapy (TRT), typically involves attaching a therapeutic radioisotope to a vector molecule, such as a peptide or a monoclonal antibody. This vector molecule is designed to bind specifically to receptors that are overexpressed on the surface of cancer cells. The resulting compound, called a radiopharmaceutical, is administered to the patient, travels through the bloodstream, and seeks out and binds to the tumor cells. The attached radioisotope then delivers a highly localized dose of cell-killing radiation directly to the cancer site.

The choice of radioisotope is critical and depends on several factors: its half-life, the type and energy of the radiation it emits, and its chemical properties. The half-life must be long enough to allow for production, transport, attachment to the vector molecule, and delivery to the tumor, but short enough to minimize long-term radiation exposure to the patient. A half-life of a few days is often ideal for therapy.

The radiation emitted by the isotope determines its therapeutic effect. The primary goal is to damage the DNA of the cancer cells, triggering apoptosis or other pathways to cell death. Different types of radiation, including beta ($\beta$) particles, alpha ($\alpha$) particles, Auger electrons, gamma ($\gamma$) rays possess distinct therapeutic properties. Beta ($\beta$) particles are electrons or positrons with a moderate range in tissue (typically 0.5 to 1.5 mm). This makes them suitable for treating larger tumors, as they can penetrate multiple cell layers and create a "crossfire" effect, killing nearby cancer cells that may not have been directly targeted by the radiopharmaceutical (Banerjee, Pillai, and Knapp 2015). Alpha ($\alpha$) particles are helium nuclei that are much heavier and have a much shorter range (10-100 µm, just a few cell diameters). However, they deposit a very large amount of energy over this short path, resulting in dense, difficult-to-repair double-strand DNA breaks \cite{eychenne_overview_2021,zimmermann_is212_2024}. This high linear energy transfer (LET) makes them extremely potent for killing isolated cancer cells or micro-metastases. Auger electrons are low-energy electrons with an extremely short range (less than a micrometer) \cite{ku_auger_2019}. To be effective, an Auger-emitting isotope must be delivered directly into the nucleus of the cancer cell, where it can cause extensive damage to the DNA at very close range. Gamma ($\gamma$) rays has low LET and find applications in imaging cancer cells. 

Many modern radiopharmaceuticals are designed as "theranostic" pairs \cite{zhang_radiopharmaceuticals_2025}. The term, delivered from the terms, "therapy" and "diagnostic," refers to using a diagnostic imaging agent to identify patients who will benefit from a specific therapy, and then using a therapeutic agent that targets the same molecular pathway. Often, the same vector molecule is labeled with two different isotopes: a gamma- or positron-emitter for diagnostic imaging (e.g., via Single Photon Emission Computed Tomography (SPECT) or Positron Emission Tomography (PET) scans), and a beta- or alpha-emitter for therapy. This allows physicians to visualize the tumor, confirm that the drug is reaching its target, and monitor the response to treatment in a highly personalized manner.

Many radioisotope drugs are available in the market, and many are in the production pipeline. A comprehensive list of radioisotopes and TRT drugs is available in the literature \cite {zhang_radiopharmaceuticals_2025}.


\subsection{\label{sec3_2}The Challenges in Radiopharmaceutical Production} 

Despite the immense promise of targeted radionuclide therapy, its widespread application is severely hampered by the limitations in different stages of its supply chain\cite{cutler_global_2021,vogel_challenges_2021,heltemes_ensuring_2020}. The production of most medical radioisotopes begins with a highly enriched stable isotope as a precursor target. This target is then irradiated in a nuclear reactor or a cyclotron to produce the desired radioisotope. Afterwards the radioisotopes are purified from the irradiated targets. Once purified the radioisotopes are transferred to a Good Manufacturing Practice (GMP) lab \cite{de_vos_good_2005,gillings_guideline_2021} to attach them with a specific vector molecule before administered to a patient. At every stage of this chain, there are challenges. 

The global supply chain for these essential radioisotopes is fragile. We discussed this in Section \ref{sec1_3}. Technologies like MAGIS and MRLIS could supplement the precursor isotope supply, making it more readily available and cheaper. Moreover, these technologies can provide highly pure stable precursor isotope targets for the radiopharmaceutical production.

Beyond the challenges of the isotope supply chain, it is worth noting the need to develop robust methods for attaching radioisotopes to vector molecules. For instance, alpha-emitters like actinium-225 \cite{bidkar_actinium-225_2024} and thorium-227 \cite{frantellizzi_targeted_2020} are generating immense excitement for their ability to deliver highly potent, targeted radiation. Still, challenges remain in production, purification, and attaching them to vector molecules. The recoil of daughter nuclei of these isotopes tends to detach them from the vector molecule and delocalize, which is not ideal for TRT. Beta-emitters, like lutetium-177($^{177}$Lu), don’t have the delocalization issue to this extent. 

Another critical challenge lies in the purification of the final radioisotope product. After transmutation, the desired radioisotope exists as a tiny fraction within the much larger bulk of the original, unreacted stable target material \cite{dash_production_2015}. It must be separated with extremely high purity. The standard method for this is radiochemistry, which uses chemical reactions to separate different elements from each other. For example, if a target of ytterbium (element 70) is irradiated to produce lutetium (element 71), the lutetium can be chemically separated from the ytterbium. This works well when the target and product are different elements \cite{dash_production_2015}. 

However, radiochemistry is powerless when the target and product are isotopes of the same element. A chemist simply cannot separate lutetium from lutetium using chemical means. This brings us to a challenge posed over a century ago by the father of nuclear physics, Ernest Rutherford.


\subsection{\label{sec3_3}Ernest Rutherford’s Challenge from 1911}

In 1911, Ernest Rutherford, then at the University of Manchester, assigned a difficult task to his young assistant, George de Hevesy. Rutherford had a sample of lead that was contaminated with a radioactive substance known as "Radium D." He challenged de Hevesy, saying, "My boy, if you are worth your salt, you separate Radium D from all that nuisance lead" \cite{levi_george_1985,hevesy_tales}. De Hevesy, a skilled chemist, tried every chemical separation technique he knew for over a year, but failed completely. The reason for his failure, which became clear only later with the development of the concept of isotopes, was that Radium D is in fact lead-210, an isotope of lead. De Hevesy was being asked to do the chemically impossible: separate lead from lead.

This story perfectly encapsulates Rutherford’s challenge: to find a method to separate radioisotopes of the same element, an alternative to radiochemistry for cases where it cannot work. One can imagine doing the separation by physical separation techniques like Calutron (EMIS), AVLIS, MAGIS, or MRLIS.  However, except for MRLIS \cite{barr_efficient_2025} all other techniques are too big to run in a hot cell for radioisotope separation. Moreover, Calutron cannot be used to separate any radioisotopes due to the poor capture efficiency. Using a laser-based isotope selection like in AVLIS, MAGIS or MRLIS, we can also preferentially excite and select a nuclear isomer (same isotope with different nuclear excitation) when we purify from an irradiated target. These aspects are also included in table \ref{table1} comparing the isotope-separating techniques. Meeting Rutherford’s challenge is not merely an academic exercise; it can allow more efficient production routes for some of the most important medical radioisotopes, particularly $^{177}$Lu.

\subsection{\label{sec3_4}Targeted Cancer Therapy with $^{177}$Lu}

$^{177}$Lu has emerged as one of the most important therapeutic radioisotopes in modern oncology, driven by the success of two FDA-approved drugs: [$^{177}$Lu]Lu-DOTA-TATE (Lutathera) \cite{hennrich_lutathera_2019} for treating neuroendocrine tumors (NETs), and [$^{177}$Lu]Lu-PSMA-617 (Pluvicto) \cite{hennrich_177lulu-psma-617_2022} for treating metastatic castration-resistant prostate cancer. With many more $^{177}$Lu-based drugs in the clinical pipeline, the demand for this isotope is surging \cite{banerjee_lutetium-177_2015,vogel_challenges_2021}.

$^{177}$Lu has nearly ideal properties for targeted therapy. It is a beta-emitter with a half-life of 6.65 days. Its beta particles have a maximum energy of about 0.5 MeV and an average range in tissue of less than 1 millimeter, making it highly effective for irradiating microtumors while sparing nearby healthy organs \cite{banerjee_lutetium-177_2015}. It also emits low-energy gamma radiation, which is not significantly therapeutic but allows for simultaneous imaging of the drug's distribution in the body using SPECT cameras, enabling personalized dosimetry and treatment monitoring.

\begin{figure*}
\includegraphics[width=1.0\textwidth]{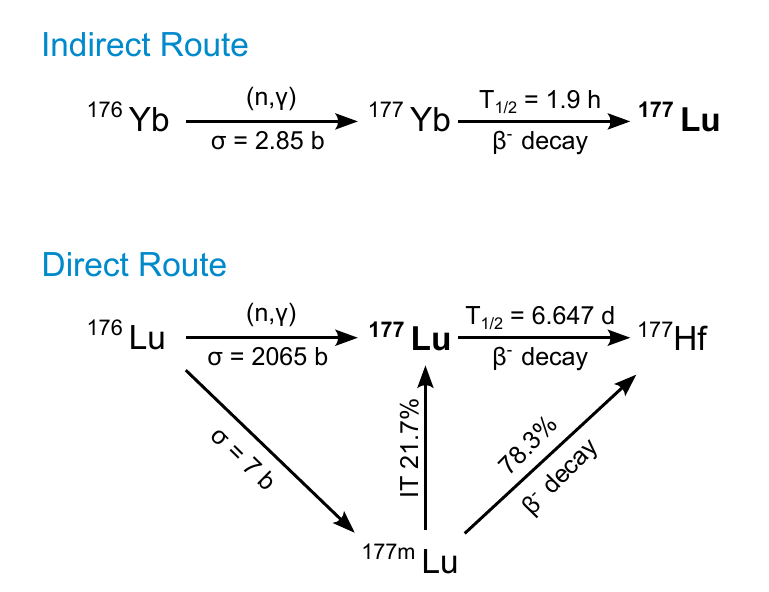}
\caption{\label{fig3} : Schematic figure showing two routes to make $^{177}$Lu }
\end{figure*}

There are two primary routes for producing $^{177}$Lu in a nuclear reactor \cite{banerjee_lutetium-177_2015, dash_production_2015}; indirect and direct routes. The schematic figure for the two routes is shown in FIG.\ref{fig3}. The indirect route starts with an enriched target of a stable isotope of ytterbium-176, $^{176}$Yb. When irradiated with neutrons, $^{176}$Yb captures a neutron to become $^{177}$Yb. $^{177}$Yb is unstable, with a short half-life of 1.9 hours, and it decays into the desired $^{177}$Lu. Because the final product (lutetium) is a different element from the target material (ytterbium), they can be separated using standard radiochemistry \cite{dash_production_2015}. This process yields what is known as "non-carrier-added" (NCA) $^{177}$Lu, which has very high specific activity (a high ratio of radioactive $^{177}$Lu to any stable lutetium isotopes), a desirable quality for radiopharmaceuticals. However, the neutron capture cross-section for $^{176}$Yb is relatively small (2.85 barns), meaning this route requires a high-flux, and therefore expensive and rare, nuclear reactor to produce sufficient quantities.

The direct route starts with an enriched target of a stable isotope of lutetium itself, $^{176}$Lu, which captures a neutron to become $^{177}$Lu. The great advantage of this route is that the neutron capture cross-section for $^{176}$Lu is enormous—2,065 barns, over 700 times larger than that of $^{176}$Yb. This means that $^{177}$Lu can be produced far more efficiently, using lower-flux, more common, and less expensive nuclear reactors, which could substantially increase global supply and lower costs.

However, the direct route has a significant drawback that has historically limited its use. The neutron irradiation of $^{176}$Lu not only produces $^{177}$Lu but also creates a long-lived nuclear isomer, $^{177m}$Lu, which has a half-life of 160 days. The presence of this long-lived, radioactive impurity is highly undesirable, as it increases the long-term radiation dose to the patient and creates a significant radioactive waste management problem. And here lies the crux of the issue: because the target ($^{176}$Lu), the product ($^{177}$Lu), and the impurity ($^{177m}$Lu) are all isotopes of the same element, they cannot be separated by radiochemistry. This is precisely Rutherford's challenge, reincarnated in a modern, critical medical context.

This is where the physical isotope separation method of MRLIS can be utilized. By offering a viable way to separate isotopes of the same element, it could allow for the use of the highly efficient direct production route for $^{177}$Lu. After irradiation of the enriched $^{176}$Lu target, the target is vaporized in vacuum with an oven/laser ablation/e-beam evaporation. The vapor is then collimated and undergoes MRLIS separation. MRLIS is estimated to take roughly an hour to separate $\sim$1 mg of $^{177}$Lu from the irradiated target. 10 $\mu$g of $^{177}$Lu can give 4 patient doses. The half-life of $^{177}$Lu is about a week. We can have an efficient pipeline to chemically incorporate the radioactive atoms into the pharmaceuticals and administer them to the patient. It is worth reiterating the high neutron capture cross-section for $^{176}$Lu, in the direct route, allowing the use of lower-flux, more common, and less expensive nuclear reactors, which could dramatically increase global supply and lower the cost. So, we believe it is economically feasible to have the MRLIS purification stage at the end of the direct route.

Achieving the direct route production of $^{177}Lu$ would represent the definitive meeting of Rutherford’s challenge and would strengthen the supply chain for one of the promising cancer-fighting agents available today.  It could also be applied to other promising radioisotopes for cancer therapy.


\subsection{\label{sec3_5}Prospects and Future Directions}

The principles applied to $^{177}$Lu extend to a wide range of other isotopes with therapeutic and diagnostic potential. Isotopes, such as mercury-197 (produced from mercury-196), strontium-85 (produced from strontium-84), and strontium-89 (produced from strontium-88) also require same-element separation and could benefit from these new physical methods. A non-exhaustive list of radioisotopes which benefits from the direct production routes and their medical uses are listed in table \ref{table2}.

\begin{longtblr}[
caption = {Radioisotopes benefits from direct isotope production routes},
  label = {table2}
                ]{colspec={X[l] X[l] X[2,l] },
                  rowhead = 1,
                  row{1} = {c, font=\bfseries},
                hlines, vlines 
                }
Radioisotope &Parent stable isotope	&Use	\\

$^{177}$Lu & $^{176}$Lu & neuroendocrine tumors\cite{hennrich_lutathera_2019}, prostate cancer\cite{hennrich_177lulu-psma-617_2022} \\

$^{99}$Mo & $^{98}$Mo or $^{100}$Mo & SPECT imaging\cite{committee_on_state_of_molybdenum-99_production_and_utilization_and_progress_toward_eliminating_use_of_highly_enriched_uranium_molybdenum-99_2016} \\

$^{197}$Hg & $^{196}$Hg & potential theranostic applications\cite{walther_theranostic_2015} \\

$^{85}$Sr & $^{84}$Sr & bone cancer\cite{charkes_early_1964} \\

$^{89}$Sr & $^{88}$Sr & bone cancer \cite{finlay_radioisotopes_2005}\\

\end{longtblr}

Despite the challenges, the effectiveness of TRT has put many radioisotope drugs in the production pipeline. A list of potential TRT drugs is available in the literature \cite {zhang_radiopharmaceuticals_2025}. Opening up direct production routes will make isotopes cheaper, high-specific-activity radioisotopes available, and make treatments more accessible.

Furthermore, the utility of isotopes in oncology is not limited to therapy. The analysis of stable isotope ratios in biological samples is emerging as a powerful tool for the early detection of cancer. Malignant tumors often exhibit altered metabolic pathways compared to healthy cells. These changes can lead to subtle but measurable shifts in the natural abundance ratios of stable isotopes like zinc, copper, and sulfur in a patient's blood or urine \cite{ploussard_urine_2010, kazi_tani_copper_2021,schilling_zinc_2021}. For example, studies have shown that the isotopic signature of metals in urine can serve as a potential biomarker for pancreatic cancer, one of the deadliest forms of the disease, offering hope for earlier and more effective diagnosis. As our ability to measure these minute isotopic variations with ultra-high sensitivity improves, this field of "isotope metallomics" promises to provide a new window into the early stages of disease, enabling intervention long before symptoms become apparent. We will discuss more about isotope metallomics in Section \ref{sec5}.


\section{\label{sec4}Prevention of Iron Deficiency: A Global Health Imperative}

Beyond the high-stakes realm of cancer therapy, isotopes play an equally vital, if less dramatic, role in addressing widespread public health crises. In this section, we will see how we can use enriched rare stable isotopes as metabolic tracers, particularly for studying iron deficiency. We will see that the high cost of isotopes and isotope detection methods is constraining this application in mitigating global health issues like iron deficiency, the most widespread nutritional disorder in the world. Its consequences, particularly when it progresses to iron deficiency anemia (IDA), are severe and far-reaching.

\subsection{\label{sec4_1}The Global Burden and Challenges of Intervention}

According to the World Health Organization (WHO), iron deficiency affects billions of people globally. It is the leading cause of anemia, which in 2023 was estimated to affect approximately 1.27 billion people, or nearly 16\% of the world's population \cite{lee_global_2025}. The highest prevalence is found in children and pregnant women in low- and middle-income countries, where diets are often dominated by plant-based staples with low iron bioavailability \cite{pasricha_iron_2021, WHO_global_anaemia_estimates}

The health consequences are profound. In children, iron is critical for neurodevelopment, and deficiency can lead to irreversible cognitive impairment, reduced school performance, and behavioral problems \cite{beard_why_2008,low_effects_2013}. It is estimated that the risk of child mortality falls by 24\% with iron supplementation \cite{scott_impact_2014}. In adults, it causes fatigue, weakness, and diminished physical work capacity, thereby impacting economic productivity \cite{haas_iron_2001}. For pregnant women, iron deficiency anemia increases the risk of premature birth, low birth weight, and maternal mortality \cite{haider_anaemia_2013}.

Addressing this public health challenge is deceptively complex. While simple iron supplementation tablets are available, their use is often plagued by poor compliance due to significant gastrointestinal side effects like constipation and nausea \cite{hyder_side-effects_2002}. A more sustainable and equitable approach is food fortification, the addition of iron to widely consumed staple foods such as wheat flour, maize flour, rice, and condiments like salt or soy sauce \cite{davidsson_bioavailability_2011,macphail_edta_1994,perez-exposito_ferrous_2005}. However, the success of any fortification program hinges on a critical parameter: bioavailability. The human body does not absorb all the iron it consumes; in fact, it absorbs only a small fraction, and this fraction varies enormously depending on the iron's chemical form and the composition of the meal \cite{perez-exposito_ferrous_2005}.

Iron in the diet exists in two forms: heme iron (from hemoglobin and myoglobin in meat, poultry, and fish) and non-heme iron (from plant-based foods and iron fortificants). Heme iron is highly bioavailable (15-35\% absorption), but non-heme iron absorption is much lower (1-15\%) and is strongly influenced by other dietary components. Inhibitors, such as phytates found in whole grains and legumes, and polyphenols found in tea and coffee, can drastically reduce non-heme iron absorption by binding to it in the gut \cite{macphail_edta_1994,davidsson_iron_1994}. Conversely, enhancers, most notably ascorbic acid (Vitamin C), can significantly boost its absorption \cite{fairweather-tait_bioavailability_1995}. Therefore, simply adding an iron compound to a food is not enough; one must ensure it is in a form that the body can actually use \cite{perez-exposito_ferrous_2005}. This is where stable isotopes of iron have become an indispensable, gold-standard tool for nutritional science.


\subsection{\label{sec4_2}Stable Isotope Tracers: The Gold Standard for Measuring Bioavailability}

To design effective fortification strategies, researchers must be able to accurately measure iron bioavailability from complex meals. Stable, non-radioactive isotopes of iron (primarily $^{54}$Fe, $^{56}$Fe, $^{57}$Fe, and $^{58}$Fe) provide a powerful and safe method to do this directly in human subjects, including vulnerable groups like children and pregnant women. The methodology, based on the principle of isotope dilution \cite{rodriguez-gonzalez_isotope_2005,abrams_using_1999}, is exquisitely precise.

A typical iron absorption study starts by taking a baseline blood sample from the subject to determine their natural iron isotope abundance pattern and their overall iron status (e.g., serum ferritin levels). The subject then consumes a test meal that has been fortified with a known quantity of a rare, enriched iron isotope, for example, $^{57}$Fe or $^{58}$Fe. This isotope acts as a tracer, or "label," for the iron in the meal. Over the next 14 days, the absorbed isotopic tracer is incorporated into the hemoglobin of newly formed red blood cells (erythrocytes). A second blood sample is drawn after the incorporation period. The next step is to determine total circulating iron in the body and measure the isotopic composition of the iron in the hemoglobin using high-precision mass spectrometry or laser spectrometry. By measuring the shift in the isotopic ratio caused by the tracer, researchers can precisely calculate the total amount of the tracer isotope that was incorporated into the red blood cells. Based on the well-established fact that approximately 80\% of absorbed iron is utilized for erythrocyte production, the total amount of iron absorbed from the meal can be determined.

This method can be made even more powerful by using a dual-isotope technique \cite{kastenmayer_double_1994}. For instance, a test meal might be labeled with $^{57}$Fe, while a reference dose of a highly absorbable iron salt (like ferrous ascorbate) labeled with a different isotope, $^{58}$Fe, is given simultaneously. By comparing the absorption of the two isotopes, researchers can determine the relative bioavailability of the iron in the food compared to the reference, controlling for individual variations in a subject's absorptive capacity.


\subsection{\label{sec4_3}High-Precision Analytical Techniques}

Measuring the tiny shifts in isotopic abundances required for these studies necessitates extremely sensitive and precise analytical instrumentation. The techniques used in this field are Thermal Ionization Mass Spectrometry (TIMS) and Multi-Collector Inductively Coupled Plasma Mass Spectrometry (MC-ICP-MS).

For a long time, Thermal Ionization Mass Spectrometry (TIMS) was the gold standard for high-precision isotope ratio measurements \cite{heumann_recent_1995}. The process involves extensive chemical purification of iron from the blood sample to remove any other elements that could interfere with the measurement. A tiny amount of the purified iron is then loaded onto a high-purity metal filament (e.g., rhenium). Inside the mass spectrometer's vacuum chamber, the filament is heated to a very high temperature ($>$1200 $^o$C), causing some of the iron atoms to be ionized. These ions are then accelerated, focused into a beam, and separated by a magnetic field according to their mass-to-charge ratio. TIMS can achieve exceptionally high precision, but it is extremely laborious and time-consuming, limiting sample throughput.

Multi-Collector Inductively Coupled Plasma Mass Spectrometry (MC-ICP-MS) needs simpler sample preparation than that for TIMS. Instead of thermal ionization, a liquid sample is introduced into an extremely hot (~6000-8000 K) argon plasma. This plasma is so energetic that it efficiently and consistently ionizes nearly all the atoms in the sample. The resulting ions are then extracted into the mass spectrometer. The key innovation of MC-ICP-MS is the use of an array of multiple ion detectors (Faraday cups) that can measure the beam intensities of all the relevant isotopes simultaneously. This simultaneous detection cancels out instabilities in the plasma source, leading to very high precision, comparable to that of TIMS, but with much simpler sample preparation and dramatically higher throughput \cite{halliday_recent_1995}. 

\begin{figure*}
\includegraphics[width=1.0\textwidth]{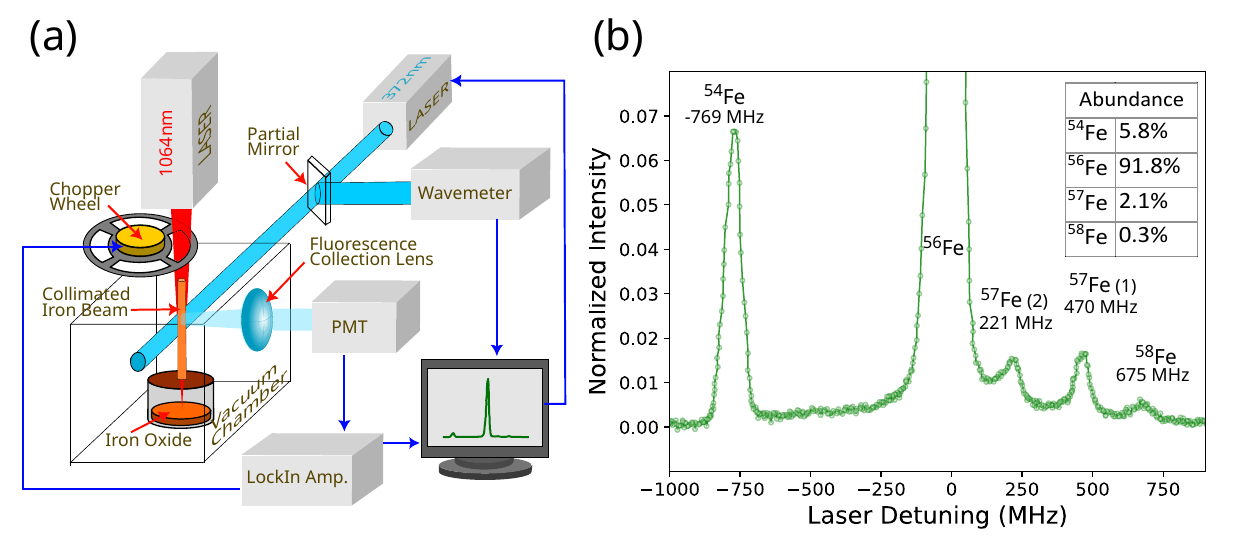}
\caption{\label{fig4} : (a) Schematic figure of a laser isotope spectroscopy setup. (b) Iron isotope spectrum measured from the setup. The full width at half maximum of all the isotope peaks are around 75 MHz. The inset table shows the natural abundance of each of the stable iron-isotopes.}
\end{figure*}

The TIMS and MC-ICP-MS remain expensive, and sample preparation is time consuming and tedious. In this context, we are developing a method using laser spectroscopy to measure isotopic abundance in biological samples. For iron isotope measurements, we use blood samples. That are oxidized in a furnace to produce a powder. The powder is then laser-ablated to generate a vapor with neutral iron atoms. The iron vapor is collimated with a slit and excited with a tunable 372 nm laser to excite an atomic transition. As we scan the laser frequency around 372 nm, each isotope of iron resonates at its unique wavelength and emits fluorescence photons. The intensity of these fluorescence photons is plotted against the laser detuning frequency to generate a spectrum with different peaks for each stable iron isotope. The peak height corresponds to the abundance of the isotope. FIG. \ref{fig4} shows the schematic of the setup and a spectrum from iron oxide powder. Iron oxide powder will be replaced by oxidized powder from blood. We can measure the isotopic abundance of other metals by choosing an appropriate laser for that metal. Previously, we employed a similar setup as an iron atom beam source to demonstrate efficient cooling of atoms through surface scattering \cite{ajith_surface_2025}. A research paper detailing laser isotope spectroscopy on blood samples is in preparation from our group.

Compared to the mass spectroscopy techniques, laser isotope spectroscopy is cheaper. Furthermore, it doesn’t need sample purification steps making the sample preparation even simpler than MC-ICP-MS. Moreover, the laser isotope spectroscopy sample preparation can be parallelized to a large number of samples. Thus, laser isotope spectroscopy is quicker and easier to do, making it scalable for large population clinical studies than was ever feasible with TIMS or MC-ICP-MS.


\subsection{\label{sec4_4}Impact on Global Health Policy and Future Outlook}

The insights gained from stable isotope studies have had a direct and profound impact on global health policies. These studies have been instrumental in validating fortificants and guiding nutritional programs. Research using stable isotopes demonstrated that sodium iron EDTA (NaFeEDTA) is a superior iron fortificant for high-phytate foods because the EDTA chelator protects the iron from being bound by phytates in the gut \cite{macphail_edta_1994,davidsson_iron_2002}. This evidence led the WHO to recommend NaFeEDTA for use in flours where phytate levels are high for adults \cite{world_health_organization_guideline_2022}. Stable isotope studies revealed why the iron status didn’t improve initially \cite{rivera_impact_2004} under the nutritional intervention by the Mexican government program PROGRESA (Oportunidades) during 1997. It was found that the reduced iron used to fortify milk-based complimentary food has low bioavailability \cite{perez-exposito_ferrous_2005}. After the study, the program used much more bioavailable iron sulfate to fortify the complementary food \cite{perez-exposito_ferrous_2005}. Stable isotope studies have been crucial for the HarvestPlus program \cite{furr_stable_2005,petry_stable_2012} and other efforts to develop biofortified crops \cite{liria-dominguez_biofortified_2023}. By measuring iron absorption from new crop varieties (e.g., high-iron beans, pearl millet, and rice) grown to naturally contain more iron, researchers can confirm that the extra iron is bioavailable and can have a real impact on iron status. Studies have precisely quantified the dose-response relationship for enhancers like Vitamin C, helping to establish the optimal amount to add to fortified foods or meals to maximize iron absorption \cite{fairweather-tait_bioavailability_1995,davidsson_iron_1994}.The bioavailability studies mentioned for iron can be extended for other mineral deficiencies as well \cite{abrams_using_1999}. Stable isotopes as tracers not only help in understanding bioavailability, but also aid in tracking the metabolism of toxic heavy metals, which is important in understanding and treating heavy metal poisoning. Similar studies are used to understand the toxicity of drugs \cite{mutlib_application_2008}.

Despite this success and wide applicability, the full potential of this technology is constrained by a two overriding factors: the immense cost of the enriched stable isotopes and limitations of mass spectrometers. At approximately \$3,500 per 10 mg for $^{58}$Fe, the cost of the tracers alone makes large-scale, population-level studies or the routine monitoring of national fortification programs prohibitively expensive. The current standard tools for isotope detection; TIMS and MC-ICP-MS are expensive, and they need extensive sample preparation for every sample, making population-level studies challenging.

Affordable, plentiful stable isotope tracers would transform public health nutrition. And the methods like laser isotope spectroscopy could make isotope detection scalable and faster. This would enable ministries of health to move from simply monitoring the iron content of fortified foods to monitoring the actual iron absorption and impact on the population. This would allow for the continuous optimization of programs, ensuring that investments in fortification lead to tangible improvements in the health, development, and economic potential of billions of people worldwide.


\section{\label{sec5}Stable Isotopes for Diagnosis of Disease: The Field of Isotope Metallomics}

In the previous section, we saw how to track metabolism by deliberately altering composition of stable isotopes. In this section, we will see how certain diseases can bring a tiny variation in isotopic composition and how this variation can be used as an early diagnostic marker. This aspect gave birth to a new field called isotope metallomics. This approach provides a dynamic window into metabolic processes that cannot be obtained by simply measuring the concentration of an element. We will see why cheaper ways to detect isotopic ratios are essential to make these early diagnostics routine and accessible.

\subsection{\label{sec5_1}The Principle of Mass-Dependent Isotopic Fractionation in Biology}

The foundation of isotope metallomics is the phenomenon of mass-dependent isotopic fractionation. Most elements with biological relevance have two or more stable isotopes (e.g., $^{40}$Ca and $^{44}$Ca; $^{63}$Cu and $^{65}$Cu; $^{64}$Zn and $^{66}$Zn). Although chemically identical, the slight mass difference between these isotopes can cause them to behave differently during physical and chemical processes \cite{urey_thermodynamic_1947}. Lighter isotopes, having a slightly higher vibrational energy and weaker chemical bonds, tend to react faster in kinetic processes \cite{bigeleisen_chemistry_1965}. They also tend to be favored in equilibrium processes like transport across cell membranes \cite{galimov_biological_1985} or incorporation into mineral lattices \cite{moynier_calcium_2017}.

As an element moves through a metabolic pathway, undergoing a series of transport, binding, and enzymatic reactions, these small fractionation effects accumulate. The result is that different tissues, organs, and biological fluids can develop distinct and reproducible isotopic signatures. A disease state that alters a metabolic pathway can therefore perturb this isotopic balance, creating a measurable shift in the isotopic composition of elements in easily accessible samples like blood or urine.

These tiny variations are seen as the changes in 4th or 5th decimal places in the relative abundance \cite{schoeller_isotope_1999}. Delta ($\delta$) notation is a convenient way to report these tiny changes. The delta value expresses the ratio of a heavy to a light isotope in a sample relative to that same ratio in a certified international standard, reported in parts per thousand (per mil, or $^{0}/_{00}$). For example, the calcium isotope composition is reported as:

\begin{equation*}
\delta^{44/42}\text{Ca}=\left[\frac{(^{44}\text{Ca}/^{42}\text{Ca} )_{sample}}{(^{44}\text{Ca}/^{42}\text{Ca} )_{standard}} -1\right]×1000
\end{equation*}

A positive $\delta$ value means the sample is enriched in the heavy isotope relative to the standard, while a negative value means it is depleted. It is these subtle $\delta$ value shifts that serve as powerful new biomarkers of metabolic function and dysfunction.


\subsection{\label{sec5_2}Osteoporosis: A Dynamic View of Bone Health with Ca and Sr Isotopes}

Osteoporosis, a disease of low bone mass and high fracture risk, represents a prime example of the diagnostic power of isotope metallomics. The current clinical gold standard for diagnosis is the DEXA scan \cite{haarbo_validation_1991, deshpande_alternatives_2023}, which measures bone mineral density. While effective, DEXA provides a static snapshot of accumulated bone loss; it cannot reveal the current rate of bone remodeling or predict future bone loss.

Stable calcium isotopes offer a solution. Bone is in a constant state of remodeling, with osteoclasts resorbing old bone and osteoblasts forming new bone \cite{hadjidakis_bone_2006}. This process fractionates calcium isotopes. During bone formation, the lighter isotopes (e.g., $^{42}$Ca) are preferentially incorporated into the bone's hydroxyapatite mineral matrix \cite{moynier_calcium_2017}. Consequently, the blood pool becomes transiently enriched in the heavier $^{44}$Ca. Conversely, during bone resorption, the lighter isotopes are released from the bone back into the blood.

This means the $\delta^{44/42}$Ca value of blood or urine acts as a direct, real-time indicator of net bone mineral balance. Net bone resorption (e.g., osteoporosis, immobilization) causes more light isotopes to be released into the blood than are being taken up by bone. The blood and urine become isotopically "light" (a lower $\delta^{44/42}$Ca value). On the other hand, the net bone formation (e.g., growth, fracture healing) causes more light isotopes sequestered into bone. The blood and urine become isotopically "heavy" (a higher $\delta^{44/42}$Ca value).

The study by Morgan et al. demonstrated the power of this technique in a study of healthy individuals confined to bed rest, a well-established model for rapid bone loss \cite{morgan_rapidly_2012}. They observed a significant decrease in the $\delta^{44/42}$Ca of the subjects' urine within just one week of immobilization, long before any change would be detectable by DEXA. Similar effects have been observed in astronauts during spaceflight, another model of extreme bone demineralization \cite{smith_bone_2005}. This makes it a promising tool for early diagnosis, for monitoring a patient's response to therapy, and for screening new potential treatments for osteoporosis. The primary barrier to its routine clinical use remains the cost of the high-precision analysis. The laser isotope spectroscopy we developed for iron-isotopes and discussed in the previous section will be well suited for reducing the cost and making these measurements routine.

The calcium metabolism can be tracked with levels of strontium \cite{vezzoli_strontium_1998,wasserman_strontium_1998}. Strontium is found to accumulate in bones \cite{cabrera_strontium_1999}. This bone-seeking nature of strontium finds application in the early diagnosis of metastatic bone cancer with radioactive $^{85}$Sr \cite{charkes_early_1964}. And the radioactive $^{89}$Sr is used for the palliative treatment of bone cancer \cite{finlay_radioisotopes_2005}. But the cost of the enriched isotopes is a main roadblock for these applications.


\subsection{\label{sec5_3}New Frontiers in Isotope-Based Diagnostics}

The principles established with calcium are now being extended to other essential metals, opening new avenues for diagnosing a host of metabolic diseases. 

Zinc is a vital cofactor for over 300 enzymes and transcription factors, playing critical roles in cell proliferation, DNA synthesis, and insulin signaling \cite{chasapis_zinc_2012}. Altered zinc metabolism is a known feature of several cancers. For instance, studies have shown that breast cancer tissue is isotopically lighter in zinc (lower $\delta^{66/64}$Zn) compared to healthy breast tissue from the same patient \cite{larner_zinc_2015}. This may reflect the high metabolic rate and zinc uptake by the tumor cells. The $\delta^{66/64}$Zn value in blood serum has been proposed as a potential non-invasive biomarker for several cancers, including breast and prostate \cite{schilling_zinc_2021, ploussard_urine_2010}. 

Copper is essential for cellular energy production (cytochrome c oxidase) and antioxidant defense (superoxide dismutase), but it is highly toxic in its free form \cite{angelova_copper_2011}. Its metabolism is therefore tightly regulated. Genetic disorders can disrupt this balance. Wilson's disease, caused by a failure to excrete excess copper, leads to toxic copper accumulation in the liver and brain. Studies have shown that patients with Wilson's disease have a distinctly lower $\delta^{65/63}$Cu value in their blood serum, providing a potential new diagnostic marker \cite{lamboux_blood_2020}. Copper also plays a key role in cancer progression, as it is a required cofactor for enzymes involved in angiogenesis (the growth of new blood vessels that feed a tumor) and metastasis. $\delta^{65/63}$Cu is a good marker for different types of cancers like, breast \cite{telouk_copper_2015}, colorectal \cite{telouk_copper_2015}, ovarian \cite{toubhans_cu_2020} and thyroid cancer \cite{kazi_tani_copper_2021} reflecting a systemic alteration of copper metabolism driven by the tumor.


\subsection{\label{sec5_4}The Technological Bottleneck and Future Promise}

The field of isotope metallomics holds immense promises for a new generation of diagnostics that probe metabolic function directly. However, its translation from research laboratory to clinical practice faces a major hurdle; the need for routine access to the expensive and complex MC-ICP-MS instrumentation required for the high-precision measurements. Early diagnosis is practical only when isotope ratio detection is accessible and low-cost. The laser isotope spectroscopy we presented in Section \ref{sec4_3} has the potential to become a tool in hospitals and clinics for this purpose.

There is a link between technologies for isotope detection and isotope purification. Both rely on similar principles and can be modified slightly to work interchangeably. For instance, isotope detection with mass spectroscopy relies on the same principles used in Calutrons for isotope separation. Similarly, the laser schemes used for enrichment can also be used for precise detection of that isotope. So, the development of technologies for isotope purification can help in the development of technologies for isotope detection.

We are currently developing technologies for enrichment and detection of stable isotopes for the applications described in Sections \ref{sec4} and \ref{sec5}. This work is being pursued at the Copenhagen Center for Biomedical Quantum Sensing, which is centered at the Niels Bohr Institute in Copenhagen. This center is funded by the Novo Nordisk Foundation in Denmark for 6 years(till 2030). A partnership will be established between the Novo Nordisk Foundation and the Pointsman Foundation to establish a new lab in Copenhagen outside the university by the year 2030.


\section{\label{sec6}Isotopes for Fundamental Physics}

The utility of isotopes extends beyond the realm of medicine and into the deepest questions of fundamental physics. The precise control and measurement of individual atoms and their nuclei provide a unique laboratory for testing the principles of quantum mechanics, particularly the long-standing debate between determinism and probability that was famously championed by Albert Einstein.


\subsection{\label{sec6_1}Einstein's Challenge: Determinism vs. Probability}

Albert Einstein, despite being one of the founding fathers of quantum theory, was deeply troubled by its implications. He famously objected to the Copenhagen Interpretation, championed by Niels Bohr, which posits that the properties of a quantum system are inherently probabilistic until a measurement is made, at which point the system's "wave function collapses" into a single, definite state. Einstein believed that this probabilistic nature was not a fundamental feature of reality but rather a reflection of our incomplete knowledge. He argued for the existence of underlying "hidden variables" that would, if known, determine the outcome of any measurement precisely. His belief in an underlying deterministic reality was famously summarized in his assertion that "God does not play dice".

This debate was sharpened by thought experiments like Erwin Schrödinger's Cat (1935), which illustrated the seemingly absurd consequences of the Copenhagen interpretation when applied to macroscopic objects. Schrödinger imagined a cat in a sealed box with a radioactive atom. If the atom decays—an event described by quantum mechanics as purely probabilistic—it triggers a mechanism that releases a poison, killing the cat. According to a strict interpretation, until the box is opened and a measurement is made, the entire system exists in a superposition of states: the atom has both decayed and not decayed, and the cat is simultaneously alive and dead.

A key experimental development in this debate was the formulation of Bell's theorem by John Bell in 1964 \cite{bell_einstein_1964} and the subsequent experiments that tested it \cite{clauser_bells_1978}. These experiments, which were recognized with the 2022 Nobel Prize in Physics, have consistently shown violations of Bell's inequalities, a result that is widely interpreted as ruling out the existence of local hidden variables \cite{aspect_experimental_1982}. This has been taken as strong evidence against Einstein's position and in favor of the inherent probabilism of quantum mechanics.

However, a subtle but crucial loophole may exist. Bell's theorem applies to elementary particles (like photons or electrons) that, as far as we know, have no internal structure. It may not necessarily apply to complex, many-body composite systems like an atomic nucleus, which consists of dozens of protons and neutrons bound by the strong force. Such a composite system possesses emergent properties that are not present in its constituent parts. One such property is the total nuclear magnetic moment, which arises from the complex interplay of the spins and orbital motions of all the nucleons. This magnetic moment is a real, physical property of the nucleus. It is not hidden; it is a "local unhidden variable". This raises a profound question: What if the seemingly random process of radioactive decay is governed by the internal evolution of such complex nuclear properties? What if we could predict when single radioactive atom will decay?


\subsection{\label{sec6_2}Trapped Ions: A Platform for Ultimate Precision}

Testing such a radical hypothesis requires measurement tools of unprecedented precision. The experimental platform of choice for this is the trapped ion. By using a combination of static and oscillating electric and magnetic fields, single charged atoms can be isolated from their environment and held nearly motionless in a vacuum for days or even months \cite{allison_hows_1987}. Two main types of traps are used: the Paul trap \cite{paul_electromagnetic_1990}, which uses an oscillating radiofrequency electric field for confinement, and the Penning trap \cite{van_dyck_high_1976}, which uses a strong static magnetic field.

Once trapped, an ion's motional and internal quantum states can be controlled with exquisite precision using lasers. Laser cooling techniques can be used to slow the ion until it is in the quantum ground state of motion, essentially reaching the lowest possible temperature allowed by quantum mechanics \cite{itano_cooling_1995,eschner_laser_2003}. With the ion held virtually still, its internal energy levels can be probed with very high accuracy. This has made trapped ions among the most accurate atomic clocks in existence \cite{dehmelt_monoion_1982}. An atomic clock's "ticking" is the frequency of an electromagnetic wave (microwave or laser light) that is perfectly resonant with a transition between two quantum states in an atom. By using long interrogation times, a technique known as Ramsey spectroscopy \cite{ramsey_experiments_1990}, the transition frequency can be measured with a fractional uncertainty of parts in $10^{18}$ or better \cite{nicholson_systematic_2015}—a precision equivalent to a clock that would not lose or gain a single second over the age of the universe.


\subsection{\label{sec6_3}A New Test of Quantum Mechanics with a Radioactive Atomic Clock}

We proposed a new test of Quantum Mechanics by making an atomic clock with a radioactive atom. The first question would be to search for a time dependent clock shift over several half-lives. This was proposed in a theoretical paper, "Do Atoms Age?" \cite{raizen_atoms_2022}, from 2022. A subsequent paper in 2025 entitled "New Tests of Quantum Mechanics with Unstable Nuclei" \cite{raizen_new_2025}, and the last paper in this conceptual trilogy was "Testing Distinguishability and Determinism in Atoms" \cite{raizen_testing_2025}. One of the experiments proposes to leverage the incredible precision of atomic clocks to search for a deterministic signal in radioactive decay. The core idea is to build an atomic clock using a radioactive isotope and search for a systematic drift in its ticking rate over several half-lives. An atomic clock's frequency is determined by the energy difference between two quantum states in an atom. This energy difference, particularly in the case of hyperfine splitting, depends sensitively on the nuclear magnetic moment. If the internal state of a radioactive nucleus evolves over time as it approaches the moment of decay, this evolution might manifest as a tiny, time-dependent change in its magnetic moment. Such a change would, in turn, cause a predictable shift in the frequency of the atomic clock. Discovering such a time-dependent shift would be revolutionary, providing the first evidence that radioactive decay is not a fundamentally random process but is instead deterministic, governed by the internal evolution of the nucleus.

The proposed candidate for this experiment is Strontium-85 ($^{85}$Sr). $^{85}$Sr is a radioactive isotope with a convenient half-life of 65 days that decays to stable Rubidium-85. It can be produced by neutron irradiation of stable $^{84}$Sr, followed by isotopic separation to create a pure sample. The experiment would involve trapping a small number of $^{85}$Sr ions in a linear Paul trap, where they form a stable "ion crystal". Using well-established laser and microwave techniques developed for ion-based atomic clocks over the past 30 years \cite{raizen_ionic_1992}, the hyperfine splitting frequency of the trapped $^{85}$Sr ions (approximately 5.67 GHz) could be measured with ultrahigh precision. By monitoring this frequency continuously over a period of months (several half-lives), one could search for any systematic drift that correlates with the age of the atoms.

This concept can be extended even further to ask whether two stable isotopes, which we assume to be perfectly identical, are truly indistinguishable. Consider the stable isotope $^{87}$Sr. Some $^{87}$Sr atoms on Earth are "old," having been produced billions of years ago during stellar nucleosynthesis. But one can also create "new" $^{87}$Sr atoms in a reactor by irradiating $^{86}$Sr. Using the separation methods developed to meet Rutherford's challenge, one could isolate these newly created $^{87}$Sr atoms. By comparing the nuclear magnetic moments (via clock frequency measurements) of the "new" and "old" atoms, one could test the fundamental assumption of indistinguishability. Current theories of nuclear structure are not precise enough to predict magnetic moments \cite{ullmann_high_2017} to the level of accuracy of atomic clocks (theory is good to about 1-10\% \cite{li_nuclear_2018}, while measurements have parts per billion precision \cite{schneider_double-trap_2017}), a difference between new and old atoms would be a paradigm-shifting discovery.

The answer to the third challenge by Einstein is unknown, but the experiments are clear;  they can be performed in the next few years and may settle the question of determinism over probability in Quantum Mechanics, a result of fundamental importance.


\section{\label{sec7}Conclusion}

Isotopes have diverse use cases spanning from practical applications in health care to answering fundamental questions on the nature of the physical world. The cost and availability of isotopes are limiting these uses and stopping the benefits from reaching more people. New technologies like MAGIS and MRLIS are poised to strengthen the isotope supply chain and lower the cost of isotopes. The realization of Maxwell's demon; Maxwell's challenge, resulted in MAGIS. MAGIS provides highly pure stable isotopes in an economic way.

Stable isotopes are used as metabolic tracers to solve global health issues like iron deficiency and as precursors for radioisotope production. The precursor stable isotope targets are irradiated to generate radioisotopes, then the radioisotopes are typically purified from the target with radiochemistry. In the case of $^{177}$Lu production, a prominent radioisotope for cancer therapy, radiochemistry restricts us to pick a production route which is not efficient. $^{177}$Lu production has a direct route that radiochemistry cannot handle, Rutherford's challenge. The new isotope separation technique MRLIS, which is compact enough to work in a hot cell, solves Rutherford's challenge and opens up direct-route production for many isotopes. For $^{177}$Lu, MRLIS will make production feasible in common small-flux irradiation facilities, thus accessible and cheaper.

Technologies for detecting isotopes are critical for using stable isotopes as metabolic tracers and measuring the isotopic composition of our bodies. The new field of isotope metallomics shows promise for detecting certain diseases at their early stages by tracking isotopic composition in blood or urine. This application will be fruitful when isotope detection is routine and cheap; laser isotope spectroscopy might be the solution here.

Finally, radioactive ion traps present a new experimental platform to test determinism and atom distinguishability. These new experiments will help solve Einstein's challenge in understanding the true nature of Quantum Mechanics. The experiments and technologies inspired by the challenges raised by Maxwell, Rutherford, and Einstein are shaping the applications of isotopes in medicine and fundamental physics. 


\begin{acknowledgments}
The authors gratefully acknowledge the support of the Copenhagen Center for Biomedical Quantum Sensing, the Pointsman Foundation and the Sid W. Richardson Foundation.
\end{acknowledgments}

\bibliography{apssamp}

\end{document}